\documentclass
[superscriptaddress,secnumarabic,amssymb,amsmath,nobibnotes,aps,prd,showkeys,showpacs,twocolumn,nofootinbib]{revtex4-1}%

\usepackage{bm}
\usepackage{mathrsfs}
\usepackage[all]{xy}
\usepackage{epsfig,subfigure}
\usepackage{latexsym, amssymb, amsmath, amsfonts}
\usepackage[english]{babel}
\usepackage[OT1]{fontenc}
\usepackage[utf8]{inputenc}
\usepackage{makeidx}
\usepackage[colorlinks=true,breaklinks=true]{hyperref}
\usepackage{graphicx, wrapfig, epsf}
\usepackage{float}
\usepackage[dvipsnames, table]{xcolor}
\hypersetup{urlcolor=BlueViolet,
	        citecolor=Plum,
	        linkcolor=OliveGreen}

\definecolor{Gray}{rgb}{.9,.9,.9}

\begin{document}

\title{Giant planet formation in Palatini gravity}

\author{Aneta Wojnar}
\email{aneta.magdalena.wojnar@ut.ee}
\affiliation{Laboratory of Theoretical Physics, Institute of Physics, University of Tartu,
W. Ostwaldi 1, 50411 Tartu, Estonia
}

\begin{abstract}
Some parts of the accretion model of the jovian planets' formation are studied in the context of Palatini gravity. We mainly focus on the critical core mass, that is, a mass for which there is no hydrostatic equilibrium solution for the planet's envelope, which is a starting point of the runaway accretion.
We also discuss the conditions needed to be satisfied by a planet such that it can posses a massive gaseous envelope around a solid core. 
\end{abstract}

\maketitle

\section{Introduction}


Since the Einstein's proposal \cite{ein1,ein2}, although being successfully tested by many observations and experiments \cite{Will:2014kxa}, and which predicted the already confirmed existence of black holes \cite{TheLIGOScientific:2017qsa,Akiyama:2019cqa,aki2,aki3,god} (see \cite{Barack:2018yly} for review), is not able to explain a certain number of cosmological and astrophysical phenomena without introducing (still) undetectable particles, many modifications to General Relativity (GR) has been proposed in order to describe processes responsible for the {\it dark side of the Universe} \cite{Copeland:2006wr,Nojiri:2006ri,nojiri2,nojiri3,Capozziello:2007ec,Carroll:2004de,cantata}. Apart from this, one would also like to address the problem of spacetime singularities \cite{Senovilla:2014gza}, to unify physics of different scales \cite{ParTom,BirDav}, as well as to explain the existence of massive compact objects exceeding theoretical predictions  \cite{lina,as,craw,NSBH,abotHBH,sak3}.

The fact that modified gravity introduces additional terms to the Poisson and hydrostatic equilibrium equations (see \cite{Saito:2015fza,olekinv,olmo_ricci}); for review, \cite{review,cantata}, which are used to describe stellar and substellar bodies, provides that our understanding of the astrophysical objects and their evolution can slightly differ when compared to the results given by Newtonian gravity. The best known examples are altered limiting masses, such as the Chandrasekhar mass for white dwarf stars \cite{Chandra,Saltas:2018mxc,Jain:2015edg,Banerjee:2017uwz,Wojnar:2020wd,Belfaqih:2021jvu,kalita,kalita2}, the minimum Main Sequence mass \cite{sak1,sak2,Crisostomi:2019yfo,gonzalo}, minimum mass for deuterium burning \cite{rosyadi}, or Jeans \cite{capJeans} and opacity mass \cite{anetaJup}. Moreover, the most prominent feature of those modifications are related to the evolution and age of the non-relativistic stars and planets: the stellar early and post-Main Sequence evolution \cite{aneta2, chow, merce, straight} or cooling processes of brown dwarfs \cite{maria} and gaseous planets \cite{anetaJup}. They also change the light elements' abundances in stellar atmospheres \cite{aneta3}.

Regarding planets, there is also a modified gravity's impact on their internal structure and evolution. Terrestrial planets, such as the Earth and other rocky planets of our Solar System, may have slightly different density profiles \cite{olek,olek2,olek3} than believed, since their interiors are obtained by the extrapolation of the PREM model \cite{prem,kus,ken,ken2,iris}, that is, their density profiles are given by the seismic data together with the assumption on hydrostatic equilibrium equation given by Newtonian physics. Moreover, since we are getting more accurate information about the internal structure of the Earth, those findings can be used to test theories of gravity \cite{olek}. Improved seismic observations \cite{mush}
provided that there is a liquid or mushy region in the inner core. It was believed that one dealt with the solid one instead. Another improvement is related to the setting of a new generation of the neutrinos' telescopes which will bring information on the matter density inside the planet, together with characteristics and abundances of light elements in the outer core, independently on a model of gravity \cite{topography,top,top2,top3}. Considering equations of state (EoS) which always carry uncertainties regarding matter properties under high pressure and temperature, behaviour of iron, being the main element of the terrestrial planets' cores, was studied in the extreme conditions of the Earth's core recreated in laboratories \cite{laser}, allowing subsequently to determine its EoS in the given pressure range in the nearest future. Therefore, all this together brings us to the exiting possibilities of testing gravity with rocky planets, such as the Earth and Mars \cite{seismic}.

Another class of planets, Jupiter-like giants, because of having many resemblances to brown dwarf stars, and since modified gravity also alters brown dwarfs' properties which can be caught by the current technology \cite{sak1,sak2,Crisostomi:2019yfo,gonzalo,rosyadi,maria}, is also expected to be used to reveal something new about various proposal of gravity in dense environments \cite{testing}. Indeed, modified gravity can have a significant effect on the late time evolution of a jovian planet and its formation processes \cite{anetaJup}. In this work we are going to focus more on the formation by a mass accretion, which is believed to be a process which happened in the case of Solar System's gaseous planets. We will demonstrate that modified gravity has an influence on the well-known scenarios of the giants' formation - although we will do so by studying a simple analytical model, the gained insight will allow us to improve the analysis by considering a more realistic assumptions in the future.

Before going to the description of some of the jovian planets' formation processes, let us just briefly introduce a model of gravity we are going to use in order to demonstrate the mentioned effect. One of the simplest case of the so-called metric-affine models of gravity is $f(\bar{R})$ gravity, which waives the assumption on the Levi-Civita connection of a metric $g$. Therefore, one deals with two independent geometric structures: the metric $g$, and the connection $\Gamma$. Such approach have a few advantages \cite{junior,sch,ol1,ol2}; one of the most prominent feature is the fact that in the vacuum the theory reduces to GR with a cosmological constant.

The action of this model is as follows:
\begin{equation}
S[g,\Gamma,\psi_m]=\frac{1}{2\kappa}\int \sqrt{-g}f(\bar{R}) d^4 x+S_{\text{matter}}[g,\psi_m],\label{action}
\end{equation}
where $\bar{R} = g^{\mu\nu}\bar{R}_{\mu\nu}(\Gamma)$ is the Palatini curvature scalar, constructed with the metric and the independent connection, while $\psi_m$ denotes as usually matter fields. The constant $\kappa=-8\pi G/c^4$; we also use $(-+++)$ signature convention. In such an approach the variation of the above action is taken with respect to the metric and to the connection. The first one provides
\begin{equation}
f'(\bar{R})\bar{R}_{\mu\nu}-\frac{1}{2}f(\bar{R})g_{\mu\nu}=\kappa T_{\mu\nu},\label{structural},
\end{equation}
with $T_{\mu\nu}=-\frac{2}{\sqrt{-g}}\frac{\delta S_m}{\delta g_{\mu\nu}}$ being the energy-momentum tensor while the prime denotes here differentiating with respect to the curvature $\bar{R}$. Contraction of the above equations with the inverse of the metric $g_{\mu\nu}$ results as  an algebraic relation between the Palatini curvature scalar and the trace of energy-momentum tensor $T$:
\begin{equation}\label{trace}
    f'(\bar{R})\bar{R}-2f(\bar{R})=\kappa T,
\end{equation}
allowing to find a relation between the scalar curvature and the trace of the energy momentum tensor. In the next sections, we will use the quadratic model
\begin{equation}
    f(\bar{R})= \bar{R}+\beta \bar{R}^2,
\end{equation}
which provides, when applied to (\ref{trace}), that $\bar{R}=-\kappa T$, such that all modifications appearing in the well-known equations are functions of baryonic matter fields.

However, the variation with respect to the independent connection introduces conformal structures, that is, from the field equation (the covariant derivative is defined by the independent connection)
\begin{equation}
\nabla_\beta(\sqrt{-g}f'(\bar{R}(T))g^{\mu\nu})=0.\label{con}
\end{equation}
one sees that there exists a metric tensor $h_{\mu\nu}$ coformally related to the metric $g_{\mu\nu}$
\begin{equation}
    h_{\mu\nu} = f'(\bar{R}(T))g_{\mu\nu}
\end{equation}
allowing to write \eqref{con} as
\begin{equation}
\nabla_\beta(\sqrt{-h}h^{\mu\nu})=0.\label{con2}
\end{equation}
Therefore, the conncetion $\Gamma$ is the Levi-Civita one of the metric $h_{\mu\nu}$. It should be also commented that it does not lead to any extra degree of freedom \cite{DeFelice:2010aj,BSS,SSB}. 

In the next section we will introduce basic notions and equations needed to the planets' description. In the further parts we will discuss conditions which are necessary for the atmosphere presence, the structure of the atmosphere-core, and the critical core mass, which is a main point in the giant planets' formation processes.

\section{Planets in modified gravity}

Let us focus now on the most crucial ingredients to describe a planet and some part of its evolution in a given theory of gravity. Therefore, the hydrostatic equilibrium equation for quadratic model in Palatini $f(R)$ gravity for non-relativistic objects is given as \cite{gonzalo} 
\begin{equation}\label{pres}
 p'=-\frac{G M(r)}{r^2}\rho\big(1+\kappa c^2 \beta [r\rho'-3\rho]\big) \ ,
\end{equation}
where prime denotes the derivative with respect to the radius coordinate $r$ in Jordan frame, the constant $\kappa=-8\pi G/c^4$ with $G$ being Newtonian constant and $c$ speed of light. Regarding the mass function, let us use
 the common definition\footnote{However, one should be concerned about the extra terms appearing there when modified gravity is considered; see, for example \cite{olek,olek2} on modified gravity issues. In Palatini gravity the extra terms were shown to be insignificant terms in the mass \cite{olek,olmo_ricci}.}
\begin{equation}\label{masa}
   M'(r)=4\pi r^2\rho(r).
\end{equation}
Introducing the surface gravity $g$: 
\begin{equation}\label{surf}
 g\equiv\frac{G M(r)}{r^2},
\end{equation}
which is usually taken as a constant on the planet's surface, the hydrostatic equation (\ref{pres}) can be written in a more convenient way 
\begin{equation}\label{hyd}
 p'=-g\rho\left( 1+8\beta\frac{g}{c^2 r} \right).
\end{equation}
The temperature gradient which describes the temperature $T$ variation with depth,
\begin{equation}
 \nabla_{\text{rad}}:=\left(\frac{d \ln{T}}{d\ln{p}}\right)_{\text{rad}},
\end{equation}
turns out to be modified in Palatini gravity \cite{aneta2}
\begin{equation}\label{grad}
  \nabla_{\text{rad}}=\frac{3\kappa_RLp}{64\pi \sigma G MT^4}\left(1+8\beta\frac{G M}{c^2 r^3}\right)^{-1},
\end{equation}
where $L$ is the local luminosity, $\sigma$ the Stefan-Boltzmann constant while $\kappa_R$ is the Rosseland mean opacity. 

On the other hand, we also need an equation which describes matter in the planet's interior. In what follows, we will not focus on the core's description\footnote{Whose interior in Palatini gravity can be describe in the same way as terrestrial planets (however, with rather higher internal pressures than the ones occuring in rocky planets), which has been already studied in Palatini gravity \cite{olek}.} but the gaseous envelope only, which in our toy model can be simply described by the ideal gas equation
\begin{equation}
    p=\frac{k_B}{\mu m_H}\rho T,
\end{equation}
where $k_B$ is the Boltzman constant, $\mu$ the mean molecular weight while $m_H$ is the mass of a hydrogen atom, and $T$ temperature.

Before considering the conditions for gaseous planets formation, let us recall some particular notions related to the formation of the gaseous giants which will be used in the further part of the text:
\begin{itemize}
    \item {\it Planetesimals}.\\ Bodies of radius $\sim10-100$ km with their orbital evolution dominated by mutual gravitational interactions, in which however aerodynamic forces between gas disk and solid particles (and they stickness) are still important.
    \item{\it Isolation mass}\\ The maximum mass of a protoplanet/core acquired by consuming all planetesimals in its neighbour. It is given by
    \begin{equation}\label{miso}
        M_\text{iso}=\frac{8}{\sqrt{3}}\pi^\frac{3}{2}C^\frac{3}{2}M_*^{-\frac{1}{2}}\Sigma^\frac{3}{2}r^3
    \end{equation}
    where $C=2\sqrt{3}$ and $\Sigma$ is a surface density, depending on the disk's model.
    \item {\it Snowline}\\
    Region in the protoplanetary disk beyond which the temperature falls below $\sim150-170$ K (above those values water is in the vapor phase at the early stage of the planet formation in low pressure). For the Solar System, snowline was located at $\sim2.7$ AU. Habitable terrestrial planets must have formed in the interior region to the snowline.
\end{itemize}

\section{Conditions for the existence of a gaseous envelope around the planet's core}

In what follows, we will assume that we are dealing with a solid core, formed via two-body collisions in the same way as in the case of the terrestrial planets \cite{armitage}. If the core grows fast enough to exceed a certain mass value before the protoplanetary disk's dissipation, the hydrostatic balance between the core and envelope is not hold anymore, causing gas accretion on to the core.
However, before analyzing this critical situation of the giant planets' formation in modified gravity, let us firstly discuss the minimum core mass (MCM) the planet/core must have in order to have {\it some} atmosphere. Later on, we will discuss conditions needed to sustain a massive envelope, that is, when its mass is considerable significant with respect to the core's one.

Before going further, we need to underline that we are dealing with a toy model, therefore the numerical output should not be treated seriously (however, it has still reasonable values when compared to the more realistic models of giant planets' formation). Our aim, similarly as in our previous work \cite{anetaJup}, is to trace the modified gravity effects in the formation and evolution of substellar objects, and to indicate a necessary modification which should be taken into account in realistic modelling in the framework of any GR extensions.

\subsection{Atmosphere presence}

Let us consider a spherical-symmetric body of the mass $M_c$ with density $\rho_m$. Then its radius is given simply as $R_s=(3M_c)^\frac{1}{3}/(4\pi\rho_m)^\frac{1}{3}$. Since in many theories of gravity the gravitational energy is slightly modified (see \cite{mof, maria}, the surface escape speed from such an object in our model is expressed by
\begin{equation}
    v_\text{esc}=\sqrt{\frac{2GM_c}{R_s}\left(1+8\beta\frac{GM_c}{c^2R_s^3} \right)},
\end{equation}
where the effects of modified gravity are taken into account.
This speed must be greater than the relative speed of the molecules moving through the gas, that is, the sound speed $c_s$, in order the planet embedded within the gas disk to be able to maintain a bound atmosphere. The sound speed in the protoplanetary disk is given by the Keplerian velocity $v_K$ and the disk thickness:

\begin{equation}\label{speed}
    c_s = \left( \frac{h}{r_d} \right) v_K,
\end{equation}
where $h$ is the vertical disc scale-height and $r_d$ is the cylindrical distance between the planet and its parent star, while 
\begin{equation}
    v_K =\sqrt{\frac{GM_*}{r}\left(1+8\beta\frac{GM_*}{c^2r^3} \right)},
\end{equation}
where $r$ is an orbital radius and $M_*$ the parent star's mass. Let us notice that for the thin disk $r\approx r_d$.

Therefore, the condition $v_\text{esc}>c_s$ provides the boundary mass for the solid core, that is, the MCM for the atmosphere presence 
\begin{equation}\label{bigenv}
    M_\text{c}>\left(\frac{3}{32\pi}\right)^\frac{1}{3}\left(\frac{h}{r}\right)^3\frac{\left(M_*\left(1+3\alpha\frac{ M_*}{\pi r^3} \right)\right)^\frac{3}{2}}{\rho_m^\frac{1}{3}r^\frac{3}{2}\left(1+4\alpha\rho_m\right)^\frac{3}{2}},
\end{equation}
where we have defined $\alpha:=8\pi G\beta/(3c^2)$ for simplicity. The above relation obviously depends on the applied theory of gravity. This mass is small, and the effect of modified gravity can be insignificant. For instance, considering an icy body $\rho_m=1$ g~cm$^{-3}$ at the distance $r=5$ AU in a disk with the thickness  $h/r_d=0.05$ surrounding a star with solar properties ($M_*=M_\odot$) gives, in the case of Newtonian gravity, $5\times10^{-4}\times M_\text{Earth}$, while for Palatini one:
\begin{equation}
     M_c\gtrsim \left\{
  \begin{array}{ll}
    4.76\times10^{-4}\times M_\text{Earth} & \text{for  } \alpha=0.03\\
     5.24\times10^{-4}\times M_\text{Earth} & \text{for  } \alpha=-0.03.
  \end{array}
\right.
\end{equation}

\subsection{Significantly massive envelope}
Since slight envelopes do not have any large effect on the dynamics of the giant planets, let us consider now a case of a MCM that is able to maintain a non-negligible fraction of the core mass. 
One says that the mass of the gaseous envelope is significant if its core-total mass fraction, denoted further by $\epsilon$, is about 0.1. In what follows, we will estimate this mass in the case of the considered model of gravity.

The hydrostatic equilibrium equation for a planet with mass $M_p$ is given by (\ref{pres})
\begin{equation}
    \frac{dp}{dr}=-\frac{GM_p}{r^2}\rho(r)\left(1+3\alpha\frac{M_p}{\pi r^3} \right),
\end{equation}
where we assume that the mass of the planet $M_p$ can be taken as a constant, while $r$ is a distance from the planet's center.  Moreover, in order to examine the problem analytically, let us assume that the envelope is also isothermal such that we can use a simple equation of state of the form
\begin{equation}\label{eos}
    p(r)=c^2_s\rho(r).
\end{equation}
Therefore, we can integrate the equation (\ref{pres}) using (\ref{eos}) to get the radial density profile of the gas in the envelope
\begin{equation}\label{radden}
    \text{ln}\rho =\frac{GM_p}{c^2_s}\frac{1}{r}
    \left( 1+\frac{3}{4}\alpha \frac{M_p}{\pi r^3} \right) +\text{const},
\end{equation}
where the constant in the above expression might be obtained by matching the envelope density (\ref{radden}) to the density in the unperturbed disk at the distance $r_\text{out}=2GM_p/{c^2_s}$ given by matching the escape velocity with the disc sound speed. Doing so, we can finally write
\begin{equation}\label{profil}
    \rho(r)=\rho_0 \text{exp}\left[ \frac{GM_p}{c^2_sr}\left( 1+\frac{3}{4}\alpha \frac{M_p}{\pi r^3} \right)-\frac{1}{2} \right],
\end{equation}
where $\rho_0$ is the density at the disk mid-plane written in terms of the vertical scale-hight $h$ and the surface density $\Sigma$, that is, (see the full description of the disk properties in, for example, \cite{armitage})
\begin{equation}
    \rho_0=\left(\frac{1}{\sqrt{2\pi}}\right)\frac{\Sigma}{h}.
\end{equation}
From the minimum mass Solar Nebula model \cite{wei} the surface density was found to be $\Sigma\sim r^{-3/2}$. In the case of solid materials which the the planet's core is made of, one deals with the following relations, depending on the snowline \cite{nebula}:
\begin{align}
    \Sigma_\text{rock}=&7.1\left(\frac{r}{1\,\text{AU}}\right)^{-3/2}\text{g cm}^{-2}\;\text{for}\;r<2.7\,\text{AU},\\
     \Sigma_\text{rock/ice}=&30\left(\frac{r}{1\,\text{AU}}\right)^{-3/2}\text{g cm}^{-2}\;\text{for}\;r>2.7\,\text{AU}.
\end{align}
The discontinuity for the solids at $r=2.7$ AU results from the fact that there is icy material in the outer disk which is destroyed in the hotter, that is, inner region, of the disk.

\begin{figure*}[h]
    \includegraphics[width=0.8\linewidth]{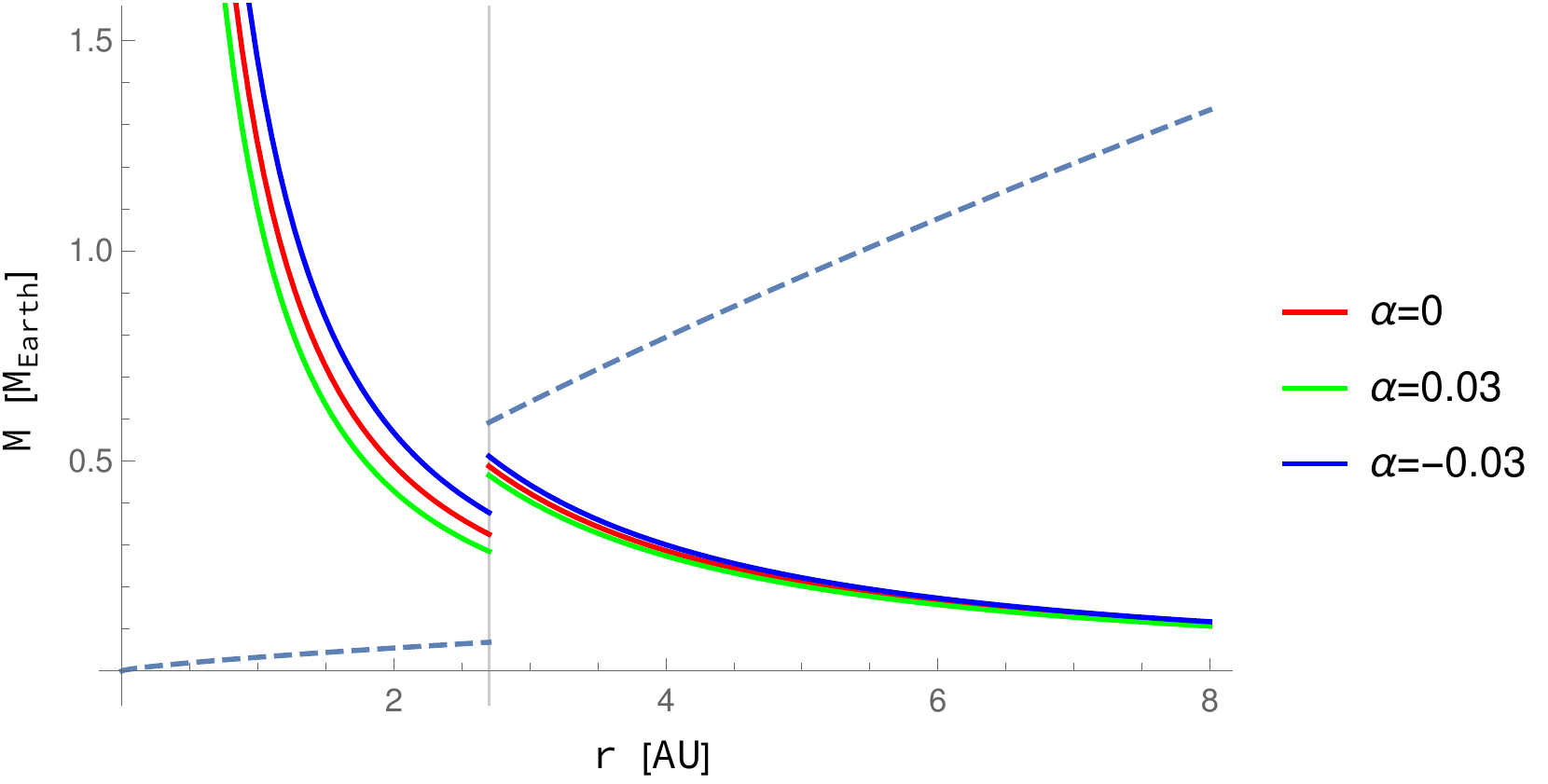}
\caption{[color online] Isolation mass (\ref{miso}) (dashed line) and the minimum planet's mass needed to sustain a massive envelope given by (\ref{bigenv}) for different values of the paremeter $\alpha$ (solid lines). The snowline at $r=2.7$ AU and other properties of the protoplanetary disk with a Solar mass star are given assuming the Hayashi minimum mass Solar Nebula model \cite{nebula}. For the interior to the snowline, $\rho_m=3$ g cm$^{-2}$, and beyond the snowline, $\rho_m=1$ g cm$^{-2}$.}
\label{fig1}
\end{figure*}

Since most of the envelope's mass with the derived density profile (\ref{profil}) lies in a shell close to the surface of the planet's solid core, the envelope mass can be then approximated as 
\begin{align}
    M_\text{env}&=\frac{4}{3}\pi R^3_s\rho(R_s)\\
    &= \frac{4}{3} R^3_s \rho_0 \text{exp}\left[ \frac{2GM_p}{c^2_sR_s}\left( 1+\frac{3}{4}\alpha \frac{M_p}{\pi R_s^3} \right)-\frac{1}{2} \right],\nonumber 
\end{align}
with $\rho(R_s)$ defined as the envelope density taken at the surface of the core $R_s$. The condition for the non-negligible fraction of the total mass of the planet we are looking for is given simply by
\begin{equation}
    M_\text{env}>\epsilon M_p,
\end{equation}
from which we derive 
\begin{equation}\label{condi}
    M_p\gtrsim \left(\frac{3}{4\pi\rho_m}\right)^\frac{1}{2}\left(\frac{c_s^2}{G}\right)^\frac{3}{2}\left(
    \frac{\mathrm{ln}\frac{\epsilon\rho_m}{\rho_0}}{
    1+\alpha\rho_m}\right)^\frac{3}{2}
\end{equation}
where we have used again $R_s=(3M_p)^\frac{1}{3}/(4\pi\rho_m)^\frac{1}{3}$. Let us notice that the value of the protoplanet's mass for which it will start to gain a considerable envelope depends on the disk sound speed, and modified gravity. The disk sound speed is a decreasing function of radius (\ref{speed}), so for GR ($\alpha=0$) protoplanets with lower masses located in the cooler regions of the disk (that is, those ones which are at larger distance from their parent star) do already acquire significant amount of gas in comparison to the ones placed closer to the star. However, as observed from the result (\ref{condi}) and figure (\ref{fig1}), modified gravity can slightly alter that scenario. One sees that an inner gaseous planet may posses a significant envelope being closer (further) to the parent star for positive (negative) parameter $\alpha$, in comparison to the Newtonian model. In the outer region, beyond the snowline, the difference decreases with the distance, approaching the well-known scenario.

\section{Core-envelope structure}
The existence of a critical core mass is a main point of the core accretion model \cite{stev}, even so simplified as in our case. In what follows, we will demonstrate that such formation models do indeed depend on a model of gravity.

By the critical core mass it is understood that crossing that value one cannot find a hydrostatic solution for the surrounding gaseous envelope. To calculate this mass, let us consider a non-rotating planet with mass $M_p$ with well defined solid core mass $M_c$ and an envelope mass $M_\text{env}$. The conservation equations for the structure of the envelope in hydrostatic and thermal equilibrium for Palatini $f(R)$ gravity are
\begin{align}
    \frac{dM}{dr}&=4\pi r^2 \rho,\\
    \frac{dp}{dr}&=-\frac{GM}{r^2}\rho\left(1+8\beta\frac{GM}{c^2r^3} \right)\label{prespl},
\end{align}
where $M$ is the total mass enclosed within radius $r$. Therefore, we do not assume anymore that the significant part of the above mass is the core one (negligible envelope's mass); that is, those relations are valid for arbitrary mass of the envelope. Moreover, the temperature varies with depth as the energy produced at core by the accretion of planetesimals is transported to the surface of the planet. It happens via radiative diffusion or convection with the first one taking place when 
\begin{equation}
     \nabla_\text{rad}\leq\nabla_\text{ad},
\end{equation}
known as a Schwarzschild criterion \cite{schw,schw2}, where the gradient describing the temperature $T$ variation with depth is given by (\ref{grad}).

To fully describe the envelope, apart from the equations (\ref{masa}), (\ref{prespl}), and the temperatue gradient (\ref{grad}), one needs expressions for the equation of state, luminosity, and Rosseland mean opacity with appropriate boundary conditions. Assuming that the luminosity is produced by the collision of planetisimals with the core provides that the the luminosity is given by ($\dot{M}_\text{core}$ is the core accretion rate)
\begin{equation}\label{lum}
    L=\frac{4}{7}\Omega\frac{GM_\text{core}\dot{M}_\text{core}}{R_s}
\end{equation}
and it is constant throughout the envelope \footnote{$\Omega$ is related to the modified gravity term \cite{artur} which in our approximation is $\approx1$.}. The inner boundary condition is simply given by $M=M_\text{core}$ at $r=R_s$ while for the outer one, since the outer radius of the planet which is supposed to match smoothly to the disk is not unambiguously defined, one has
\begin{equation}
    r_\text{out}=\text{min}(r_{acc},r_H),
\end{equation}
where the accretion radius $r_{acc}=\frac{GM_p}{c_s^2}$ measures the maximum distance for which the gas moving with sound speed $c_s$ in the disk is bound to the planet, while the Hill sphere radius 
$r_H=\left( \frac{M_p}{3M_*}r \right)^\frac{1}{3}$ is a measure of the distance at which shear present
in the Keplerian disk\footnote{It is a disk surrounding a massive body such as a star, made of material which obeys the Keplerian laws of motion. Let us notice that Kepler's laws can be modified in some theories of gravity \cite{zhao}.} unbinds gas from the envelope of the planet. Since the gas in the envelope must be bound to the planet, the outer radius must be smaller than any of those two distances. Then, for $r=r_\text{out}$ we have that $M=M_p$ while pressure and temperature take the values of the disk ones: $p_\text{disk}$ and $T_\text{disk}$, respectively.

\subsection{Critical core mass}

Let us assume now that the envelope is fully radiative. This assumption will allow us to study accretion analytically and to trace the effects of modified gravity incorporated in jovian planets' formation. Physically it means that the planet's core is separated from the protoplanetary disc by the radiative envelope, and because of that the core can be treated as independent of the disk's pressure and temperature \cite{stev}. In more realistic case envelopes of massive planets have convective and radiative zones, therefore modelling them requires numerical treatment. It should be however noticed that because the Schwarzschild criterion, which determines if one deals with radiative or convective energy transport through the planet's medium, depends on theory of gravity, therefore the results obtained from numerical analysis and simulations will differ with respect to the models which are based on Newtonian/GR equations.

The envelope being in the thermodynamical equilibrium means that its temperature is smoothly matched to the temperature of the gas in the protoplanetary disk: the excess heat produced in the core by bombarding planetisemals fades with time. In such a situation the density profile is given by (\ref{profil}). However, if the core is slowly increasing its mass, the envelope's mass also grows, altering at the same time the core-total mass fraction. Further growing envelope's mass reaches a point in this process when it cannot be supported by the increasing pressure near the core. In what follows we are going to find a critical core mass for which the massive envelope is not supported anymore by the core's conditions. Therefore, we will find an approximated solution in the form of the density profile within a radiative envelope. Integrating it will allow to obtain the mass of the envelope, and subsequently, the mass of the critical core as a function of total planet's mass, opacity, and accretion rate. 

\begin{figure*}[h]
    \includegraphics[width=0.7\linewidth]{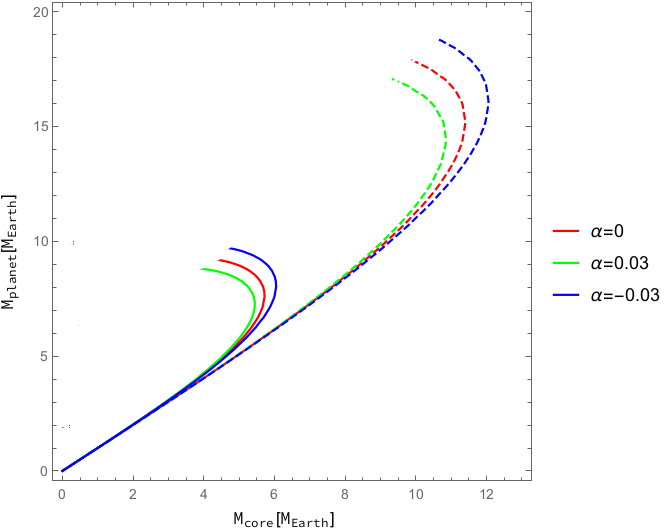}
\caption{[color online] The planet's core mass $M_\text{core}$ dependence on the total mass $M_p$ given by the equation (\ref{rozw}) for a constant opacity and two constant core accretion rates. The dashed curves are given by the five-fold rate with respect to the solid ones.}
\label{fig2}
\end{figure*}

Starting with (\ref{grad}) and (\ref{prespl}) we write
\begin{equation}
    \frac{\partial T}{\partial p}= 
    \frac{3\kappa_R L}{64\pi \sigma GMT^3}\left( 1+8\beta \frac{GM}{c^2 r^3} \right)^{-1},
\end{equation}
which can be integrated from the outer boundary inward. We will assume that $M(r)\approx M_p$ while the luminosity $L$ and the Rosseland opacity $\kappa_R$ are also constants: 
\begin{equation}
   \int^T_{T_\text{disk}} T^3 dT= 
    \frac{3\kappa_R L}{64\pi \sigma GM_p\left( 1+8\beta \frac{GM_p}{c^2 r^3} \right)}\int^p_{p_\text{disk}}dp.
\end{equation}
Since in the planet's interior we deal with $T^4>>T^4_\text{disk}$ and $p>>p_\text{disk}$, the integration of the above expression with these approximations yields
\begin{equation}
    T^4\simeq  \frac{3\kappa_R L p}{64\pi \sigma GM_p}\left( 1+8\beta \frac{GM_p}{c^2 r^3} \right)^{-1}.
\end{equation}
Applying to it the ideal gas equation of state
\begin{equation}
    p=\frac{k_B}{\mu m_p}\rho T
\end{equation}
together with the temperature stratification equation
\begin{equation}\label{strat}
    \frac{dT}{dr} = \frac{3\kappa_R L\rho}{64\pi \sigma r^2 T^3}
\end{equation}
to get rid of $T^3$, and integrating it again with respect to radius provides
\begin{equation}
    T\simeq \frac{\mu m_p}{k_B}\frac{GM_p}{4r}\left( 1+8\beta \frac{GM_p}{c^2 r^3} \right),
\end{equation}
while the density profile within the envelope is given as
\begin{equation}
    \rho(r)=\frac{64\pi \sigma}{3\kappa_R L}\left( \frac{\mu M_p G m_p}{4k_B} \right)^4
    \left( 1+8\beta \frac{GM_p}{c^2 r^3} \right)^4 \frac{1}{r^3}.
\end{equation}
The mass of the envelope is then obtained in the usual way:
\begin{align}
    &M_\text{env}=\int_{R_s}^{r_\text{out}}4\pi r^2 \rho(r) dr\\
    &=\frac{256\pi^2\sigma}{3\kappa_RL}\left( \frac{\mu m_p GM_p}{k_B} \right)^4
    \left[ \text{ln}\frac{r_\text{out}}{R_s}+\frac{8\beta GM_p}{c^2}\left(\frac{1}{R^3_s}-\frac{1}{r^3_\text{out}}\right) \right].\nonumber
\end{align}
The above equation, apart from the obvious dependence on modified gravity represented by the $\beta$-term, is strongly dependent on the planet's mass and its core (via the luminosity expression (\ref{lum}) with the radius $R_s$ given by $M_\text{core}=4\pi R^3_s\rho_\text{core}/3$). Since $M_\text{core}=M_p-M_\text{env}$, one has
\begin{equation}\label{rozw}
    M_\text{core}=M_p - \frac{C_1}{\kappa_R \dot{M}_\text{core}}\frac{M_p^4}{M^{2/3}_\text{core}}\left[ 
     \text{ln}\frac{r_\text{out}}{R_s} +\alpha C_2 M_p\right] 
\end{equation}
where, using $\alpha:=8\pi G\beta/(3c^3)$, 
\begin{align}
    C_1=& \frac{7}{3}\frac{64\pi^2\sigma G^3}{(4\pi\rho_\text{core}/3)^\frac{1}{3}}\left(\frac{\mu m_p}{k_B}\right)^4 \\
    C_2=& \frac{3}{\pi}\left(\frac{1}{R^3_s}-\frac{1}{r^3_\text{out}}\right).
\end{align}
Let us consider now a simplified case, when the accretion rate $\dot M_\text{core}$ is a constant. Despite this, the equation (\ref{rozw}) does not have any real solutions for $M_p$ with large $M_\text{core}$. We can also see it from our approximated solutions given by the figure \ref{fig2}, obtained by taking the opacity $\kappa_R=1$ with a few values of the parameter $\alpha$. 

\section{Conclusions} 

Previous studies on the late evolution of jovian planets and brown dwarfs \cite{maria,anetaJup} pointed out that cooling processes of those objects are impacted by the slight modifications of the structural equations - even so small as the ones introduced by Palatini $f(R)$ gravity. Since a similar set of equations, such as, for instance, the hydrostatic equilibrium equations and Schwarzschield criterion, are also used in simulations of giant planets' formation processes, thus one also expects to find differences with respect to the well-known scenarios, which are obtained by assuming Newtonian gravity. Because of that fact we wanted to analyse and to trace the modifications of the Palatini quadratic model by the use of a simple and crude model of the jovian planets' formation \cite{stev}. Although our results can be trusted only up to the obtained order, they were however sufficient to reveal the main differences with respect to the Newtonian model, making the future studies along these lines relevant and promising. Therefore, improving the description for a more realistic case is the first step of our future research related to the physics of exoplanets in modified gravity.

The results presented in our work can be summarized into three main points:
\begin{itemize}
    \item \textbf{Existance of the atmosphere}\\  Even very small astrophysical bodies, with the masses of order $10^{-4} M_\text{Earth}$, may have some atmospheres, since the gravitational attraction is strong enough to bound gas from the proto-planetary disk. Palatini $f(R)$ gravity slightly changes the MCM - the differences with respect to the Newtonian gravity are around $\pm5\%$ for our simplified model, and it can be difficult, if not impossible, even after improving the description, to use this result to test gravity models\footnote{Assuming, that one will be able to measure somehow atmosphere mass of some rocks in just formed planetary system, not mentioning the evaporation processes and solar winds which decrease the atmosphere mass with time.}.
    \item \textbf{Significantly massive atmosphere}\\
 A planet is considered to have a significantly massive atmosphere when its core takes about $10\%$ of the total planet's mass. The minimum mass a planet needs to have in order to sustain such a massive atmosphere depends on the snowline and the proto-planetary disc's properties in Newtonian gravity. It is depicted in the figure \ref{fig1} for a few values of the parameter $\alpha$, together with the isolation mass (\ref{miso}) and marked snowline. In general, one rather expects to find a planet with massive envelope beyond the snowline, since planets with masses below $0.5$ of the Earth's mass can already sustain them. On the other hand, objects with orbits placed much closer to their parent stars are required to have much larger masses. As expected, modified gravity introduces additional terms, and, depending on the sign of the theory's parameter, one can find a planet with a given mass closer to, or further from the star, in comparison to Newtonian gravity based models. This result could shed some light on the hot jupiters' formation processes {\it in situ} \cite{struve,mayoz}, although the biggest problem related to that issue is a formation of a core massive enough so close to a parent star in the {\it in situ} accretion model \cite{rebekah}. We have not studied it in this work. However, we see from the figure \ref{fig1} that modified gravity allows lower-mass planets with massive envelopes to be placed closer to the star than in the Newtonian scenario.\\
 Considering the outer region of the disk, one deals with much smaller differences with respect to the results provided by the Newtonian model, approaching the well-known scenario with increasing distance from the star. It is not surprising because in Palatini gravity the modifications are functions of density, as it can be seen in (\ref{condi}), and since in the outer region of the planetary disk the material (ice) is less dense than in the case of the inner zone (rocks), the modifications are also less significant.
 \item {\bf Critical core mass}\\
 Our simple model of accretion confirmed that early evolution of Jovian planets are also affected by modified gravity, and can change our understanding of the planets' and Solar System's formation processes. The crucial point during the formation is the so-called critical core mass. Written as a function of the total planet's mass $M_p$ and core accretion rate $\dot{M}_\text{core}$, the critical core mass is a boundary mass for which the massive envelope starts not being in equilibrium anymore; that is, the core's conditions are not sufficient to support the envelope. Before the core reaches the critical mass, the core and envelope grow, with the latter being still in a hydrostatic equilibrium since the energy released by the bombarding planetesimals and slowly contracting envelope is transported by radiative processes, as assumed in our model. Exceeding the critical core mass calculated here, given by (\ref{rozw}), results as uncontrolled accretion which terminates as soon as the gas supply is over, followed by a cooling process with still ongoing gravitational contraction, described in the case of Palatini model of gravity in \cite{anetaJup}.\\
 The approximated solutions of (\ref{rozw}) are given by the figure \ref{fig2} for two different core accretion rates and for a few values of the parameter $\alpha$, with $\kappa_R=1$ \cite{stev}. As observed, there is a maximal mass of the core beyond which hydrostatic solutions do not exist for a given core accretion rate. Higher masses are achieved for higher rates (the dashed curves are given by $5$ times higher rates with respect to the solid ones). As already mentioned, modified gravity provides a different scenario; depending on the sign, one deals with lower or higher critical core masses, with total planet's mass also slightly shifted. As a result, we may expect that giant planets may posses more or less massive cores than postulated on the base of Newtonian/GR gravity - a feature possible to measure and confront with the observational data \cite{juno,juno2,juno3}, since a mean moment of inertia in approximation depends only on the planet's total and core masses. The problem arises when accuracy is discussed: for a core with mass $10M_\text{Earth}$ the effect of such a core on the moment of inertia is about $0.1\%$ in the case of Newtonian models \cite{stev_juno}. But yet, the extended Juno mission is still ongoing, providing new data to be analyze \cite{helled}.
Additionally, with a big enough observational sample of giant exoplanets, high-resolution observations of other planetary systems' forming, the results presented here and in the previous works \cite{maria, anetaJup} can be used to statistically constrain a given model of gravity \cite{adam}.
\end{itemize}

Let us notice that similar results are expected to appear in the framework of any theory of gravity which modifies the structural equations. This can be also true even if only the mass expression differs with respect to GR one, as for example by considering a dark matter contribution \cite{axion}, especially in the late stages of the object's evolution \cite{horowitz,leane}. Definitely more studies on internal properties and evolution of jovian planets are required, especially in the light on the ongoing and future missions, providing more data on our Solar System's giants and exoplanets. Our next steps are related to improving the theoretical/numerical description of those objects. So far, we could understand how modified gravity affects early and late evolution of Jupiter-like planets by studying analytical models, which are very useful before introducing modifications in numerical analysis or simulations. In the future works we are going to study more realistic description, for example by taking into account a non-constant accretion rate and considering better suited equations of state to describe the jovian planets' interiors.

\vspace{5mm}
\noindent \textbf{Acknowledgement.} 
This work was supported by the EU through the European Regional Development Fund CoE program TK133 "The Dark Side of the Universe." The author would like to thank the members of the McWilliams Center for Cosmology from Department of Physics,  Carnegie Mellon University, Pittsburgh, USA, for their hospitality while finalizing this paper.

\end{document}